\documentclass[conference,compressed,twocolumn]{IEEEtran}

\usepackage[pdftex]{graphicx}
\usepackage{color,colortbl,multicol,amssymb,amsfonts,array}
\usepackage[cmex10]{amsmath}
\usepackage{gensymb,textcomp}
\usepackage[ansinew,utf8]{inputenc}
\usepackage[acronym,shortcuts,nomain,nopostdot]{glossaries}

\usepackage{booktabs}
\usepackage{dcolumn}

\newcolumntype{d}{D{.}{.}{-1}}
\newcolumntype{L}[1]{>{\raggedright\let\newline\\\arraybackslash\hspace{0pt}}m{#1}}
\newcolumntype{C}[1]{>{\centering\let\newline\\\arraybackslash\hspace{0pt}}m{#1}}
\newcolumntype{R}[1]{>{\raggedleft\let\newline\\\arraybackslash\hspace{0pt}}m{#1}}

\definecolor{to}{rgb}{0.8,0.8,0.8}
\definecolor{l1}{rgb}{0.95,0.95,0.95}
\definecolor{l2}{rgb}{0.90,0.90,0.90}

\newcommand{\lo}{\cellcolor{l1}}
\newcommand{\lt}{\cellcolor{l2}}
\newcommand{\tp}{\cellcolor{to}}
\newcommand{\tbt}[1]{\multicolumn{1}{c}{#1}}
\newcommand{\na}{\multicolumn{1}{c}{\textcolor[rgb]{0.75,0.75,0.75}{N/A}}}

\newcommand{\lona}{\multicolumn{1}{c}{\textcolor[rgb]{0.75,0.75,0.75}{\cellcolor{l1}N/A}}}

\newcommand{\ze}{\multicolumn{1}{d}{\color[rgb]{0.75,0.75,0.75}0}}
\newcommand{\zeg}{\multicolumn{1}{d}{\lt\color[rgb]{0.75,0.75,0.75}0}}
\newcommand{\zel}{\multicolumn{1}{d|}{\color[rgb]{0.75,0.75,0.75}0}}
\newcommand{\zegl}{\multicolumn{1}{d|}{\lt\color[rgb]{0.75,0.75,0.75}0}}
\newcommand{\ve}{\multicolumn{1}{d}{\color[rgb]{0.75,0.75,0.75}20}}
\newcommand{\vel}{\multicolumn{1}{d|}{\color[rgb]{0.75,0.75,0.75}20}}

\definecolor{hyp}{rgb}{0,0,0.3} 
\usepackage[colorlinks=true,linkcolor=hyp,filecolor=hyp,citecolor=hyp,urlcolor=hyp]{hyperref}

\newglossarystyle{acro}{
    \glossarystyle{super}

}
\newcommand{\acro}[2]{\newacronym{#1}{#1}{#2}}      
\newcommand{\acrol}[3]{\newacronym{#1}{#2}{#3}}     

\newcommand{\acrop}[4]{\newacronym[plural=#3,firstplural=#4 (#3)]{#1}{#1}{#2}}

\newcommand{\acF}[1]{\acf{#1}\glsunset{#1}}

\acro{3GPP}{3rd generation partnership project}
\acrol{1-D}{\mbox{1-D}}{one-dimensional}
\acrol{2-D}{\mbox{2-D}}{two-dimensional}
\acrol{3-D}{\mbox{3-D}}{three-dimensional}
\acrol{4-D}{\mbox{4-D}}{four-dimensional}
\acrol{6-D}{\mbox{6-D}}{six-dimensional}
\acro{3G}{third generation}
\acro{4G}{fourth generation}
\acro{5G}{fifth generation}

\acrop{AoA}{azimuth angle of arrival}{AoAs}{azimuth angles of arrival}
\acrop{AoD}{azimuth angle of departure}{AoDs}{azimuth angles of departure}
\acro{ASA}{azimuth spread of arrival}
\acro{AS}{angular spread}
\acro{ASD}{azimuth spread of departure}
\acro{AWGN}{additive white Gaussian noise}
\acro{AGC}{adaptive gain control}
\acro{ACF}{autocorrelation function}
\acro{ASE}{average squared error}

\acro{BD}{block diagonalization}
\acro{BS}{base station}
\acro{BC}{broadcast channel}
\acro{BPSK}{binary phase-shift keying}
\acro{BP}{break point}

\acro{CDF}{cumulative distribution function}
\acro{CIR}{channel impulse response}
\acro{CoMP}{coordinated multi-point}
\acro{CoF}{cost of feedback}
\acro{CRONOS}{cellular radio network simulator}
\acro{CSI}{channel state information}
\acro{COST}{European Cooperation in Science and Technology}
\acro{CPS}{cyber physical system}

\acro{DS}{delay spread}
\acro{DD}{Doppler-delay}
\acro{DFT}{discrete Fourier transform}
\acro{DPC}{dirty-paper coding}
\acro{D2D}{device-to-device}

\acro{ESA}{elevation spread of arrival}
\acro{ESD}{elevation spread of departure}
\acrop{EoA}{elevation angle of arrival}{EoAs}{elevation angles of arrival}
\acrop{EoD}{elevation angle of departure}{EoDs}{elevation angles of departure}

\acro{FBR}{feedback rate}
\acro{FBS}{first-bounce scatterer}
\acro{FI}{feedback interval}
\acro{FR}{frequency response}
\acro{FR1}{frequency reuse 1}
\acro{FR4}{frequency reuse 4}
\acro{FIR}{finite impulse response}
\acro{FFT}{fast Fourier transform}
\acro{FWHM}{full width at half maximum}
\acro{FDD}{frequency-division duplexing}

\acro{GCS}{global coordinate system}
\acro{GF}{geometry factor}
\acro{GR}{ground reflection}
\acro{GoC}{gain of CoMP}
\acro{GoP}{gain of prediction}
\acro{GPS}{global positioning system}
\acro{GSCM}{geometry-based stochastic channel model}

\acro{HHI}{Heinrich Hertz Institute}
\acro{HPBW}{half-power beamwidth}

\acro{ISD}{inter site distance}
\acrol{iid}{i.i.d.}{independent and identically distributed}
\acro{IDFT}{inverse discrete Fourier transform}
\acro{IFFT}{inverse fast Fourier transform}
\acro{IR}{impulse response}
\acro{ITU}{International Telecommunication Union}
\acro{IF}{intermediate frequency}

\acro{JT}{joint transmission}
\acro{JCF}{joint correlation function}

\acro{KF}{Ricean K-factor}
\acro{KPI}{key performance indicator}

\acro{LBS}{last-bounce scatterer}
\acro{LHCP}{left hand circular polarized}
\acro{LOS}{line of sight}
\acro{LSP}{large-scale parameter}
\acro{LTE}{long term evolution}
\acro{LSAS}{large-scale antenna systems}
\acro{LSF}{large-scale fading}

\acro{MET}{multi-user eigenmode transmission}
\acro{MIMO}{multiple-input multiple-output}
\acro{MISO}{multiple-input single-output}
\acro{MMSE}{minimum mean square error}
\acro{MPC}{multipath component}
\acro{MS}{mobile station}
\acro{MSE}{mean square error}
\acro{MT}{mobile terminal}
\acro{MU-MIMO}{multi-user MIMO}
\acro{MIMOSA}{MIMO over satellite}
\acro{MAC}{multiple-access channel}
\acro{MRT}{maximum ratio transmission}

\acro{NLOS}{non-line of sight}
\acro{NR}{new radio}

\acro{OFDM}{orthogonal frequency division multiplexing}
\acro{O2I}{outdoor-to-indoor}
\acro{ODA}{omni directional array}

\acro{PDP}{power delay profile}
\acro{PDF}{probability density function}
\acro{PG}{path gain}
\acro{PL}{path loss}
\acro{PUCA}{polarized uniform circular array}
\acro{PAS}{power-angular spectrum}
\acro{P2P}{peer-to-peer}
\acro{PSCS}{Propsound channel sounder}

\acro{QAM}{quadrature amplitude modulation}
\acro{QuaDRiGa}{quasi deterministic radio channel generator}
\acro{QRDRLS}{QR-decomposition recursive least squares}
\acro{QoS}{quality of service}

\acro{RB}{resource block}
\acro{RHCP}{right hand circular polarized}
\acro{RLS}{recursive least-squares}
\acro{RLE}{run-length encoding}
\acro{RX}{receiver}
\acro{RMS}{root mean square}
\acro{RAN}{radio access network}
\acro{RSRP}{reference signal received power}
\acro{RF}{radio frequency}

\acro{SAGE}{space-alternating generalized expectation-maximization}
\acro{SCM}{spatial channel model}
\acro{SF}{shadow fading}
\acro{SV}{singular value}
\acro{SVD}{singular value decomposition}
\acro{SNR}{signal to noise ratio}
\acro{SIR}{signal to interference ratio}
\acro{SINR}{signal to interference and noise ratio}
\acro{SISO}{single input single output}
\acro{SIMO}{single input multiple output}
\acro{STD}{standard deviation}
\acro{SSG}{state sequence generator}
\acro{SSF}{small-scale-fading}
\acro{SOS}{sum-of-sinusoids}

\acro{TP}{throughput}
\acro{TUB}{Technische Universit\"{a}t Berlin}
\acro{TX}{transmitter}

\acro{ULA}{uniform linear array}
\acro{UML}{unified modeling language}
\acro{UMi}{urban-microcell}
\acro{UMa}{urban-macrocell}
\acro{UAV}{unmanned aerial vehicle}

\acro{VR}{visibility region}

\acro{WINNER}{Wireless World Initiative for New Radio}
\acro{WSSUS}{wide sense stationary uncorrelated scattering}
\acro{WSS}{wide-sense stationary}
\acro{WGS}{world geodetic system}

\acro{XPD}{cross-polarization discrimination}
\acro{XPR}{cross polarization ratio}

\acro{ZF}{zero-forcing}
\acro{ZoA}{zenith angle of arrival}
\acro{ZoD}{zenith angle of departure}
\acro{ZSA}{zenith angle spread of arrival}
\acro{ZSD}{zenith angle spread of departure}
\acro{ZCR}{zero-crossing rate}

\makeatother
\begin{document}

\author{
    \IEEEauthorblockN{Stephan Jaeckel\IEEEauthorrefmark{1}, Nick Turay\IEEEauthorrefmark{1}, Leszek Raschkowski\IEEEauthorrefmark{1}, Lars Thiele\IEEEauthorrefmark{1}, Risto Vuohtoniemi\IEEEauthorrefmark{3}, Marko Sonkki\IEEEauthorrefmark{3}}
    \IEEEauthorblockN{Veikko Hovinen\IEEEauthorrefmark{3}, Frank Burkhardt\IEEEauthorrefmark{2}, Prasanth Karunakaran\IEEEauthorrefmark{2}, and Thomas Heyn\IEEEauthorrefmark{2}}
    \IEEEauthorblockA{\IEEEauthorrefmark{1} Fraunhofer Heinrich Hertz Institute, Berlin, Germany, stephan.jaeckel@hhi.fraunhofer.de}
    \IEEEauthorblockA{\IEEEauthorrefmark{3} Centre for Wireless Communications, Oulu, Finnland}
    \IEEEauthorblockA{\IEEEauthorrefmark{2} Fraunhofer Institute for Integrated Circuits, Erlangen, Germany}
}

\title{Industrial Indoor Measurements from 2-6 GHz for the 3GPP-NR and QuaDRiGa Channel Model}

\maketitle

\begin{abstract}
Providing reliable low latency wireless links for advanced manufacturing and processing systems is a vision of Industry 4.0. Developing, testing and rating requires accurate models of the radio propagation channel. The current 3GPP-NR model as well as the QuaDRiGa model lack the propagation parameters for the industrial indoor scenario. To close this gap, measurements were conducted at 2.37~GHz and 5.4~GHz at operational Siemens premises in Nuremberg, Germany. Furthermore, the campaign was planned to allow the test and parameterization of new features of the QuaDRiGa channel model such as support for \ac{D2D} radio links and spatial consistency. A total of 5.9~km measurement track was used to extract the statistical model parameters for \ac{LOS} and Non-LOS propagation conditions. It was found that the metallic walls and objects in the halls create a rich scattering environment, where a large number of multipath components arrive at the receiver from all directions. This leads to a robust communication link, provided that the transceivers can handle the interference. The extracted parameters can be used in geometric-stochastic channel models such as QuaDRiGa to support simulation studies, both on link and system level.
\end{abstract}

\section{Introduction}

The tremendous developments in the fields of electronics, communication, and advanced manufacturing in the past years are leading to a shift from digital to intelligent production methods in manufacturing companies \cite{8207346}. This modernization in manufacturing is based on the integration of new technologies into a cyber-physical system commonly known as Industry~4.0. The aim of Industry~4.0 is to establish a highly flexible production model of customized products and services where realtime interactions between humans, products, and devices during the production process are feasible \cite{7382284}. One technology that is often discussed in this context is \ac{D2D} communication. It offers new ways to communicate with less latency, transfer large amounts of data, and reduce the load in the core network \cite{6805125}.

\Acp{GSCM} such as the \ac{3GPP} \ac{NR} model \cite{3gpp_tr_38901_v1500} and the QuaDRiGa channel model \cite{quadriga_www} offer efficient ways to evaluate the performance of mobile wireless communication systems during standardization and before product development stages. These models emulate the wireless propagation environment by clearly defined stochastic processes, but the interactions of \acp{TX} and \acp{RX} with this randomized environment are purely deterministic and predictable. This approach drastically reduces the complexity compared to, e.g., ray tracing, but it requires parameterization to emulate the wireless propagation environment correctly. Finding the model parameters is a labor intensive and time consuming task. One approach is to use fully deterministic models such as ray tracing to determine the \ac{GSCM} parameters. Another way is to perform channel sounding measurements in a real environment and then extract the model parameters from the recorded \acp{CIR}. This has been done for many outdoor (e.g., \cite{Jaeckel2013,Raschkowski2016}) and some office type scenarios (e.g., \cite{Peter2016a}). So far, the \ac{3GPP} community has focussed on outdoor environments and some office type scenarios \cite{3gpp_tr_38901_v1500}. Additional extensions cover vehicle-to-everything \cite{3gpp_tr_37885_v1520}, non-terrestrial networks \cite{3gpp_tr_38811_v1500}, and aerial vehicles \cite{3gpp_tr_36777_v1500}. However, the influence of an industrial production environment on the radio transmission, both for \ac{D2D} as well as infrastructure-based systems, is not well understood and the corresponding \ac{GSCM} parameters are missing. To close this gap, we performed radio channel measurements in five different factory halls at operational Siemens premises in Nuremberg, Germany. The measurements were done at 2.37~GHz and 5.4~GHz, covered a total of 5.9~km indoor measurement tracks, and led to more than 5,000 data-points for all relevant parameters. The results allow the use of the \ac{3GPP} and QuaDRiGa channel models for industrial systems in a frequency range from 2~GHz to 6~GHz and thus enable more accurate simulation studies for \acf{5G} and beyond-5G wireless communications systems. This is especially relevant for so-called private campus networks which will be deployed at around 3.5~GHz in many countries.

\begin{figure*}
    \centering
    \footnotesize
    \includegraphics[width=180mm]{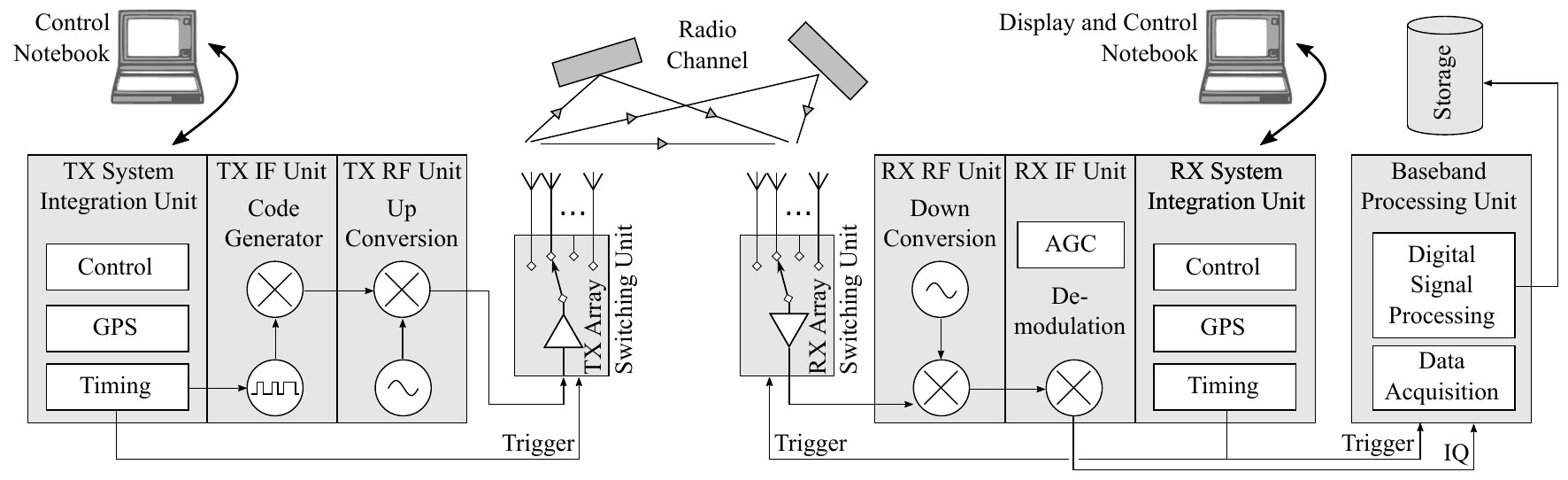}
    \vspace{-\baselineskip}
    \caption{Propsound channel sounder architecture}
    \label{fig:sounder}
\end{figure*}

\section{Measurements}\label{sec:ind_meas_measurements}

\begin{figure}[p]
    \centering
    \footnotesize
    \includegraphics[width=89mm]{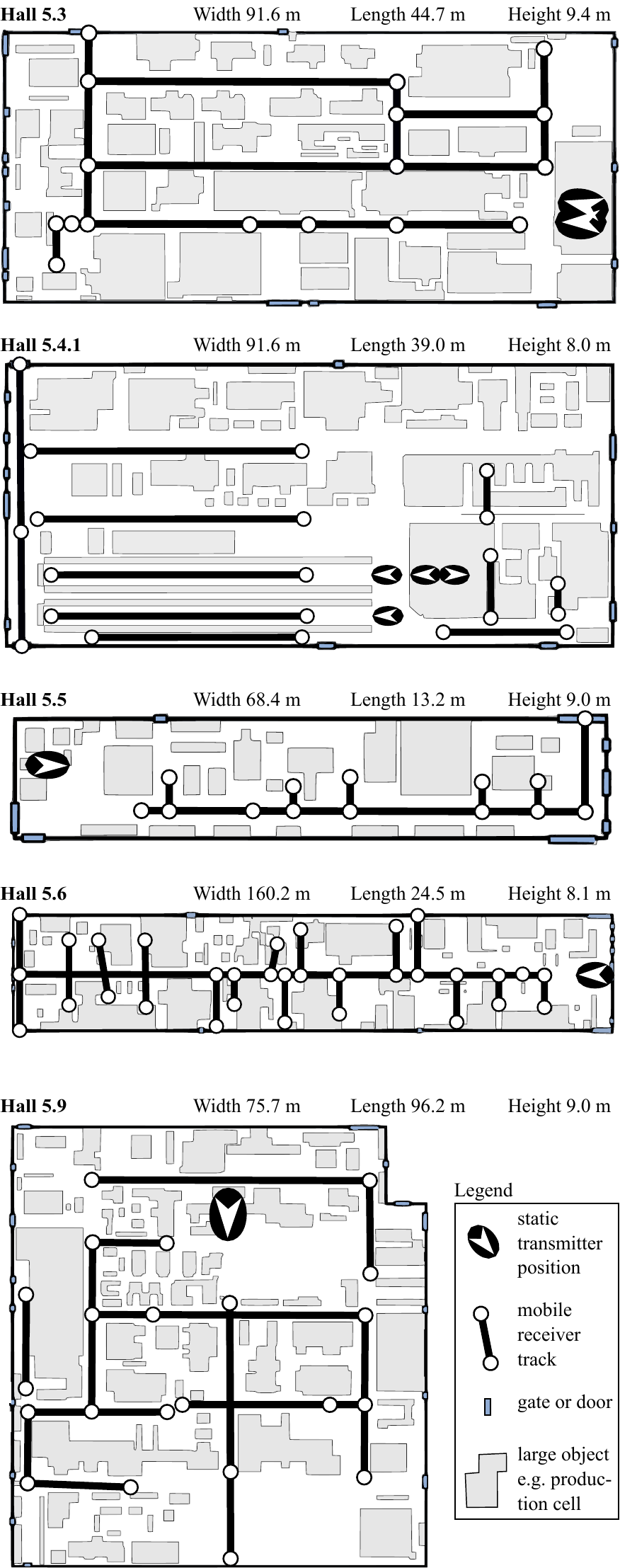}
    \vspace{-\baselineskip}
    \caption{Floor plans of the five measurement halls}
    \label{fig:Measurementhalls}
\end{figure}

Measurements were done in an industrial indoor environment within five different factory halls in Nuremberg, Germany. The halls show different properties with regard to geometry, structure and installed objects. This leads to different propagation conditions and a different behavior of the signal transmission within each hall. Several measurement tracks were defined along which a transceiver unit was moved accordingly, as depicted in Figure~\ref{fig:Measurementhalls}. An arrow symbolizes the \ac{TX} position in the single mobility scenario and the black lines indicate the measurement tracks. Start and end points are represented by a white circle. The exact position information of the \ac{TX} and \ac{RX} during the recording was measured using a laser ranger and a distance wheel. The \ac{RX} antenna height was always set to approx.\ 2~m whereas the \ac{TX} height was varied from 2~m to 8~m to cover different deployment scenarios. For example, two machines might communicate with each other when they are moving around in the factory. In this dual mobility scenario, both are at a height of 2~m. The fixed communication infrastructure, on the other hand, might be installed under the roof of the factory at a height of 6~m or 8~m. The majority of measurements were done at a variable distance between the \ac{TX} and the \ac{RX} by either moving the \ac{RX} (single mobility) or both the \ac{RX} and the \ac{TX} units (dual mobility). To study the influence of the \ac{LOS} propagation, roughly 1/3 of the measurements were done in \ac{LOS} conditions and 2/3 were in \ac{NLOS} conditions. The measurements were done at distances ranging from 5~m to 150~m for both \ac{LOS} and \ac{NLOS} channels. A total of 208 tracks were measured, resulting in a total of 5.9~km measurement tracks. An overview of the measurement parameters is given in Table~\ref{tab: Measurement Parameter}.

\begin{table}[b]
\setlength{\tabcolsep}{2pt}
\centering
\scriptsize
\caption{Measurement Parameters}\label{tab: Measurement Parameter}
\begin{tabular}{l | C{18mm} | C{18mm} | C{18mm}}
Parameter           & Setup 1 & Setup 2 & Setup 3 \\
\hline\hline
Center Frequency    & \multicolumn{2}{c|}{5.4~GHz} & 2.37~GHz \\
Bandwidth           & \multicolumn{2}{c|}{200~MHz} & 50~MHz\\
TX power            & \multicolumn{2}{c|}{23~dBm}  & 13~dBm   \\
Noise floor         & \multicolumn{2}{c|}{-101.3~dBm}  &  -102.7~dBm   \\
Max. CIR length     & \multicolumn{2}{c|}{ 4.2~\textmu s}  &  4.6~\textmu s   \\
\cmidrule{1-4}
TX Antenna Config.  & ODA & Cross & Planar \\
No. TX elements     & 32 & 30 & 32 \\
TX element gain     & 7.9~dBi   & 8.1~dBi & 8.2~dBi  \\
TX array aperture   & 205\degree & 90\degree & 90\degree \\
\cmidrule{1-4}
RX Antenna Config.  & \multicolumn{2}{c|}{ODA} & ODA \\
No. RX elements     & \multicolumn{2}{c|}{50} & 56 \\
RX element gain     & \multicolumn{2}{c|}{8.0~dBi} & 7.2~dBi \\
RX array aperture   & \multicolumn{2}{c|}{360\degree} & 360\degree \\
\end{tabular}
\end{table}

A wideband Propsound channel sounder was used to capture the in-phase and quadrature (IQ) data. To avoid interfering the factory processes, and to ensure an interference free data capture, the \ac{RF} spectrum was constantly screened for interference. In addition, the transceiver system was calibrated before and after the measurements were carried out to minimize the systems influence on the captured data. The architecture of the Propsound channel sounder is depicted in Figure~\ref{fig:sounder}. Both the \ac{TX} and the \ac{RX} unit use an \ac{IF} of 1.45~GHz which is connected to customized \ac{RF} units for different frequency bands. The \ac{TX} generates a wideband pseudo random \ac{BPSK} signal using a direct sequence spread spectrum approach. The \ac{BPSK} signal is modulated onto the desired carrier frequency in the \ac{TX} unit and down converted back to the \ac{IF} at the \ac{RX}. A system integration unit contains all control functions, timing, synchronization, and interfaces to other units.

\begin{figure}[t]
    \centering
    \includegraphics[width=84mm]{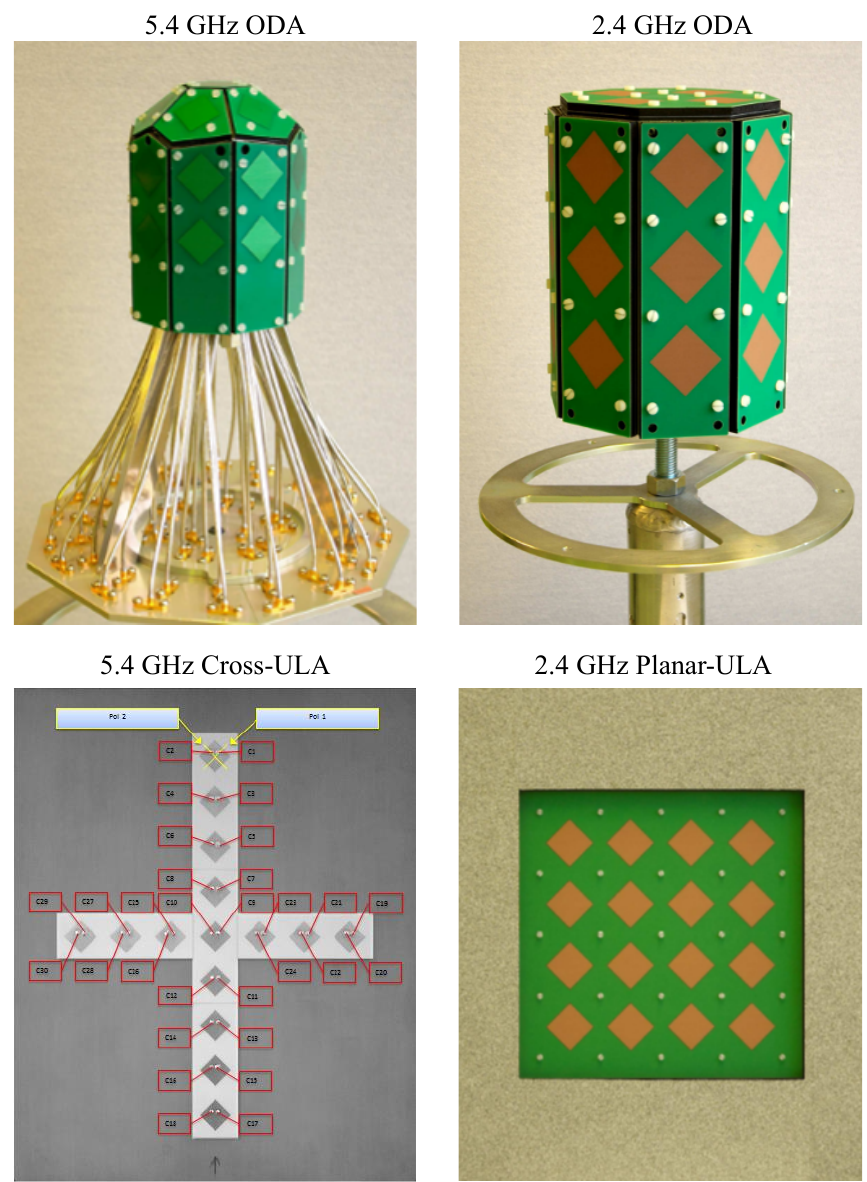}
    \vspace{-0.5\baselineskip}
    \caption{Measurement antennas}
    \label{fig:antenna_arrays}
\end{figure}

Array antennas are used to extract information about the spatial signal properties (e.g., the angles of departure and arrival). The antenna elements are sequentially switched to measure the whole \ac{MIMO} channel matrix. Hence, only a single \ac{TX}-\ac{RX} antenna pair is active at any given time. The \ac{TX} array element selection is held constant while the \ac{RX} switches through all available elements. When the \ac{RX} cycle is complete, the \ac{TX} switches to the next element and the \ac{RX} antenna switching cycle repeats. The array antennas are depicted in Figure~\ref{fig:antenna_arrays}. They are constructed from dual-polarized patch elements with a half power beamwidth of 70\degree . The polarization angles are $\pm$45\degree\ with respect to the vertical which helps to preserve the signal strength of the final $\theta$ (vertical) and $\phi$ (horizontal) polarizations in manmade environments dominated by horizontal and vertical surfaces. All antennas were calibrated via measurements in an anechoic chamber. The characterization of the antenna patterns is mandatory to correctly estimate the arrival and departure angles. The measurement settings and additional auxiliary information are time tagged and embedded in the recorded data. These are later used for post processing.

The measurements were done at two different carrier frequencies: 5.4~GHz (200~MHz bandwidth) and 2.37~GHz (50~MHz bandwidth). In both cases, the bandwidth is sufficient to extract the channel model parameters using high resolution algorithms such as \cite{Fleury1999}. To be able to interpolate between the two frequencies with regards to radio propagation behaviour, the measurement conditions for both frequency were kept very similar, e.g., the same measurement tracks and transmitter heights were used during the recordings. However, there are differences in the antenna configurations used depending on the frequency. This leads to some restrictions regarding data analysis. The three measurement setups are as follows:

\paragraph*{Setup 1 - 5.4~GHz, TX-ODA, RX-ODA configuration}
This configuration was used when both the \ac{TX} and the \ac{RX} were mobile. The \ac{TX} was equipped with an \ac{ODA} consisting of 16 dual-polarized elements covering an azimuth range of 205\degree\ . This is sufficient to calculate the azimuth angle spreads at the transmitter. The \ac{RX} used a 25-element dual-polarized \ac{ODA} that covered the whole sphere, i.e., it is possible to resolve paths from all directions.

\paragraph*{Setup 2 - 5.4~GHz TX-Cross-ULA, RX-ODA configuration}
For scenarios that included a stationary transmitter, the \ac{TX} used an \ac{ULA} with 15 dual-polarized elements arranged in a planar (cross shaped) fashion. The antenna can resolve departure angles within $\pm$45\degree\ relative to the antenna broadside. This is sufficient for the elevation direction, where there are no significant scatterers directly above or below the antenna. During the measurement planning, it was assumed that no significant paths would arrive from the rear of the antenna. However, during the data analysis it was found that the metal hall creates a rich scattering environment where the direction of many propagation paths exceeded the 90\degree\ range. This had to be considered during the data analysis.

\paragraph*{Setup 3 - 2.37 GHz TX-Planar-ULA, RX-ODA configuration}
A planar \ac{TX} antenna arrangement was used for all measurements at 2.37~GHz. Here the \ac{TX} array consists of 16 dual-polarized elements and the \ac{RX} array uses 28 dual-polarized elements. The same limitations as for the 5.4~GHz \ac{TX}-Cross-\ac{ULA} apply, i.e., the departure angles can be estimated within $\pm$45\degree\ relative to the antenna broadside. However, results from Setup~3 are comparable to Setup~2 since the measurements were done using identical transceiver positions and very similar antenna configurations.

During the measurements, the complete \ac{MIMO} channel, i.e., the broadband channel response of all \ac{TX}-\ac{RX} antenna pairs, was sampled every few centimeters. One such characterization of the channel is called a \emph{snapshot}. The first post processing step is to derive the \emph{path parameters}, i.e., the path powers, delays and angles of the \acp{MPC} from the channel snapshots. The path parameters (delay, power, angles, and polarization) are estimated from the raw measurement data by an estimation method similar to the SAGE algorithm (Space Alternating Generalized Expectation Maximization) \cite{Fleury1999}. However, the procedure used in this paper is divided into two steps. First, the path delay and the \ac{MIMO} coefficient matrix are calculated for each \ac{MPC}. This step is independent of the antennas. Then, in a second step, the departure and arrival angles are calculated. The algorithm is described in \cite{Jaeckel2017_diss}. The next step is to calculate the \acp{LSP}. The complete list of parameters consists of
\begin{itemize}
    \item the \acF{PL} and \acF{SF},
    \item the \acF{KF},
    \item the \acF{RMS} \acF{DS},
    \item the \ac{RMS} \acF{ASA},
    \item the \ac{RMS} \acF{ASD},
    \item the \ac{RMS} \acF{ESA},
    \item the \ac{RMS} \acF{ESD}, and
    \item the \acF{XPR}.
\end{itemize}
\Ac{SSF}, fast fluctuations of the power of a \ac{MPC}, can lead to strong fluctuations of these parameters even in subsequent snapshots. For example, different \ac{DS} values might be calculated from two successive snapshots, even if they were measured only centimeters away from each other. To reduce this effect, Jalden \emph{et al.} \cite{Jalden2007a} suggested to average the results within a certain radius. Here, this \emph{averaging interval} was set to 1~m for a good balance between the reduction of \ac{SSF} effects and the number of measurement samples available for further analysis. For example, a typical measurement track has a length of about 60~m. With one snapshot per wavelength, a total of 700 snapshots is captured for this track. The track is split into 60 \emph{averaging intervals}, each containing 12 snapshots. The \acp{LSP} are calculated for each snapshot and are averaged within the interval - leading to 60 values for the \ac{DS}, the \ac{KF}, the \ac{SF}, and so on. In this way, 5,468 sample points were obtained from the whole measurements. The dependency of the parameters on the \ac{TX}-\ac{RX} distance, the \ac{TX} height, and the carrier frequency was evaluated. The results of this evaluation are presented and discussed in the following section.

\section{Results and Discussion}
Based on the distance, \ac{TX} height, and frequency, there are eight variables that describe the distribution of a \ac{LSP}. These variables are
\begin{enumerate}
  \item the reference value $\mu$ (\emph{mu}) at 1~GHz, 1~m distance, and 1~m TX height,
  \item the reference \ac{STD} $\sigma$ (\emph{sigma}) at 1~GHz, 1~m distance, and 1~m TX height,
  \item the decorrelation distance $\lambda$ (\emph{lambda}) in meters,
  \item the frequency dependence $\gamma$ (\emph{gamma}) of the reference value scaling with $\log_{10}(f_\text{GHz})$,
  \item the distance dependence $\epsilon$ (\emph{epsilon}) of the reference value scaling with $\log_{10}(d_\text{2D})$,
  \item the height dependence $\zeta$ (\emph{zeta}) of the reference value scaling with $\log_{10}(h_\text{TX})$,
  \item the frequency dependence $\delta$ (\emph{delta}) of the reference \ac{STD} scaling with $\log_{10}(f_\text{GHz})$,
  \item the distance dependence $\kappa$ (\emph{kappa}) of the reference \ac{STD} scaling with $\log_{10}(d_\text{2D})$.
\end{enumerate}
These parameters are used in \acp{GSCM} such as the QuaDRiGa model or the 3GPP-NR model \cite{3gpp_tr_38901_v1500} to generate artificial channel coefficients for simulation studies. The values $V$ of a \ac{LSP} are calculated by
\begin{multline}\label{eq:lsf_ds}
    V =
        {V}_\mu +
        {V}_\gamma \cdot  \log_{10} f_\text{GHz}  +
        {V}_\epsilon \cdot  \log_{10} d_\text{2D}  +
        {V}_\zeta \cdot  \log_{10} h_\text{BS} +\\
 {X} \left( {V}_\sigma +
        {V}_\delta \cdot \log_{10} f_\text{GHz}  + \right.
        \left. {V}_\kappa \cdot \log_{10} d_\text{2D} \right)\text{,}
\end{multline}
where ${X}$ is a Normal distributed random variables having zero-mean and unit variance. Likewise, the measurements are analyzed by fitting this formula to the more than 5,000 data points. The measurement tracks were separated into \ac{LOS} and \ac{NLOS} and the parameters were fitted for the three measurement setups independently as well as for the combined dataset\footnote{The combined parameters in Table~\ref{tab:all_parameters} were obtained by fitting \eqref{eq:lsf_ds} to a combination of datapoints from the measurements, not averaging the values from the individual setups.}. The analysis results are shown in Table~\ref{tab:all_parameters}. As a reference, the last columns in the table contain the parameters of the popular 3GPP 38.901 indoor office scenario \cite{3gpp_tr_38901_v1500}. These values also have been used in industrial settings in the past due to the lack of consolidated parameters for the industrial indoor scenario. In the following, the analysis results are discussed.

\begin{table*}
    \setlength{\tabcolsep}{2pt}
    \scriptsize
    \centering
    \caption{Industrial Large-Scale Parameters}\label{tab:all_parameters}
\begin{tabular}{ llc | dd | dd | dd | dd | dd }

Parameter & Unit &
    & \multicolumn{2}{c|}{Setup 1}
    & \multicolumn{2}{c|}{Setup 2}
    & \multicolumn{2}{c|}{Setup 3}
    & \multicolumn{2}{c|}{\lt Combined}
    & \multicolumn{2}{c}{3GPP 38.901}\\

&&& \multicolumn{2}{c|}{5.4 GHz, TX-ODA}
    & \multicolumn{2}{c|}{5.4~GHz TX-Cross}
    & \multicolumn{2}{c|}{2.37~GHz TX-Planar}
    & \multicolumn{2}{c|}{\lt Indoor Industrial}
    & \multicolumn{2}{c}{Indoor Office}\\

&&& \multicolumn{1}{C{11mm}}{LOS} & \multicolumn{1}{C{11mm}|}{NLOS}
    & \multicolumn{1}{C{11mm}}{LOS} & \multicolumn{1}{C{11mm}|}{NLOS}
    & \multicolumn{1}{C{11mm}}{LOS} & \multicolumn{1}{C{11mm}|}{NLOS}
    & \multicolumn{1}{C{11mm}}{\lt LOS} & \multicolumn{1}{C{11mm}|}{\lt NLOS}
    & \multicolumn{1}{C{11mm}}{LOS} & \multicolumn{1}{C{11mm}}{NLOS} \\

\hline\hline
\textbf{PL}        & dB                      & $\text{PL}_\mu$     &  36.1 & 34.4 & 34.3 & 39.2 & 37.2 & 30.1 & \lt 36.3 & \lt 29.1 & 32.4 & 17.3 \\
Shadow Fading      & dB                      & $\text{PL}_\sigma$  &   1.6 & 1.6 & 1.5 & 3.1 & 1.7 & 1.7 & \lt 1.8 & \lt 1.15 & 3.0 & 8.0  \\
SF decorr. dist.   & m                       & $\text{SF}_\lambda$ &  20.6 & 8.0 & 3.5 & 34.2 & 14.8 & 32.1 & \lt 15.0 & \lt 30.0 & 10.0 & 6.0 \\
PL freq. dep.      & dB/log10(GHz)           & $\text{PL}_\gamma$  & \ve & \vel & \ve & \vel & \ve & \vel & \lt 19.5 & \lt 25.4 & 20.0 & 24.9 \\
PL dist. dep.      & dB/log10(m)             & $\text{PL}_\epsilon$&  18.5 & 21.7 & 19.3 & 21.7 & 17.6 & 24.7 & \lt 18.3 & \lt 24.1 & 17.3 & 38.3 \\
SF freq. dep.      & dB/log10(GHz)           & $\text{SF}_\delta$  & \ze & \zel & \ze & \zel & \ze & \zel & \lt -0.3 & \lt 3.15 & \ze & \ze \\
\cmidrule{1-13}
\textbf{DS}        & log10(s)                & $\text{DS}_\mu$     & -7.22 & -7.12 & -7.47 & -7.19 & -7.78 & -7.72 & \lt -8.3 & \lt -8.19 & -7.69 & -7.17 \\
DS STD             & log10(s)                & $\text{DS}_\sigma$  & 0.08 & 0.08 & 0.14 & 0.09 & 0.11 & 0.12 & \lt 0.09 & \lt 0.11 & 0.18 & 0.055 \\
DS decorr. dist.   & m                       & $\text{DS}_\lambda$ & 42.1 & 36.4 & 24.2 & 10.5 & 99.7 & 127.2 & \lt 50.0 & \lt 52.0 & 8.0 & 5.0 \\
DS freq. dep.      & log10(s)/log10(GHz)     & $\text{DS}_\gamma$  & \ze & \zel & \ze & \zel & \ze & \zel & \lt 1.26 & \lt 1.37 & -0.01 & -0.28 \\
DS height dep.     & log10(s)/log10(m)       & $\text{DS}_\zeta$   & 0.36 & 0.31 & 0.56 & 0.17 & 0.4 & 0.38 & \lt 0.49 & \lt 0.3 & \ze & \ze \\
DS STD freq. dep.  & log10(s)/log10(GHz)     & $\text{DS}_\delta$  & \ze & \zel & \ze & \zel & \ze & \zel & \lt 0.07 & \zegl & \ze & 0.1  \\
Delay Factor       &                         & $r_\text{DS}$       & 2.93 & 3.03 & 2.56 & 2.89 & 2.55 & 2.74 & \lt 2.7 & \lt 3.0 & 3.6 & 3.0 \\
\cmidrule{1-13}
\textbf{KF}        & dB                      & $\text{KF}_\mu$     & -1.6 & -4.0 & 4.8 & -4.4 & 2.8 & -1.0 & \lt 7.8 & \lt 4.2 & 7.0 & \na \\
KF STD             & dB                      & $\text{KF}_\sigma$  & 2.7 & 2.5 & 2.9 & 2.4 & 2.6 & 1.9 & \lt 1.8 & \lt 1.1 & 4.0 & \na \\
KF decorr. dist.   & m                       & $\text{KF}_\lambda$ & 17.5 & 11.4 & 13.5 & 6.1 & 30.3 & 17.0 & \lt 32.0 & \lt 14.0 & 4.0 & \na \\
KF freq. dep.      & dB/log10(GHz)           & $\text{KF}_\gamma$  & \ze & \zel & \ze & \zel & \ze & \zel & \lt -7.3 & \lt -11.7 & \ze & \na \\
KF height dep.     & dB/log10(m)             & $\text{KF}_\zeta$   & -5.7 & -6.3 & -8.2 & -1.0 & -3.6 & -1.9 & \lt -7.7 & \lt -3.3 & \ze & \na \\
KF STD freq. dep.  & dB/log10(GHz)           & $\text{KF}_\delta$  & \ze & \zel & \ze & \zel & \ze & \zel & \lt 2.6 & \lt 2.2 & \ze & \na \\
\cmidrule{1-13}
\textbf{ASA}       & log10(\degree)          & $\text{ASA}_\mu$    & 1.67 & 1.61 & 1.71 & 1.61 & 1.67 & 1.56 & \lt 1.69 & \lt 1.62 & 1.781 & 1.863 \\
ASA STD            & log10(\degree)          & $\text{ASA}_\sigma$ & 0.15 & 0.18 & 0.12 & 0.2 & 0.19 & 0.26 & \lt 0.15 & \lt 0.22 & 0.119 & 0.059 \\
ASA decorr. dist.  & m                       & $\text{ASA}_\lambda$& 6.1 & 6.8 & 12.1 & 9.5 & 6.3 & 9.8 & \lt 10.0 & \lt 13.0 & 5.0 & 3.0 \\
ASA freq. dep.     & log10(\degree)/log10(GHz) & $\text{ASA}_\gamma$ & \ze & \zel & \ze & \zel & \ze & \zel & \zeg & \zegl & -0.19 & -0.11 \\
ASA STD freq. dep. & log10(\degree)/log10(GHz) & $\text{ASA}_\delta$ & \ze & \zel & \ze & \zel & \ze & \zel & \zeg & \zegl & 0.12 & 0.12 \\
\cmidrule{1-13}
\textbf{ASD}       & log10(\degree)          & $\text{ASD}_\mu$    & 1.54 & 1.64 & -0.06 & 0.14 & 0.56 & 0.75 & \lt 1.66 & \lt 1.68 & 1.60 & 1.62 \\
ASD STD            & log10(\degree)          & $\text{ASD}_\sigma$ & 0.1 & 0.1 & 0.31 & 0.29 & 0.32 & 0.24 & \lt 0.12 & \lt 0.1 & 0.18 & 0.25 \\
ASD decorr. dist.  & m                       & $\text{ASD}_\lambda$& 12.8 & 16.4 & 18.1 & 18.2 & 23.8 & 17.9 & \lt 10.0 & \lt 13.0 & 7.0 & 3.0 \\
ASD height dep.    & log10(\degree)/log10(m) & $\text{ASD}_\zeta$  & 0.05 & -0.24 & 1.02 & 1.43 & 0.22 & 0.29 & \lt 0.1 & \lt -0.2 &  \ze & \ze \\
\cmidrule{1-13}
\textbf{ESA}       & log10(\degree)          & $\text{ESA}_\mu$    & 1.61 & 1.72 & 1.69 & 1.19 & 1.71 & 1.82 & \lt 1.64 & \lt 1.64 & 1.44 & 1.387\\
ESA STD            & log10(\degree)          & $\text{ESA}_\sigma$ & 0.07 & -0.11 & -0.1 & 0.25 & 0.09 & -0.12 & \lt 0.01 & \lt 0.06 & 0.264 & 0.746 \\
ESA decorr. dist.  & m                       & $\text{ESA}_\lambda$& 13.3 & 11.4 & 6.4 & 12.9 & 10.3 & 21.1 & \lt 10.0 & \lt 20.0 & 4.0 & 4.0 \\
ESA freq. dep.     & log10(\degree)/log10(GHz) & $\text{ESA}_\gamma$ & \ze & \zel & \ze & \zel & \ze & \zel & \zeg & \zegl & -0.26 & -0.15 \\
ESA dist. dep.     & log10(\degree)/log10(m) & $\text{ESA}_\epsilon$& -0.5 & -0.54 & -0.53 & -0.34 & -0.54 & -0.65 & \lt -0.5 & \lt -0.5 & \ze & \ze \\
ESA STD freq. dep. & log10(\degree)/log10(GHz) & $\text{ESA}_\delta$ & \ze & \zel & \ze & \zel & \ze & \zel & \zeg & \zegl & -0.04 & -0.09 \\
ESA STD dist. dep. & log10(\degree)/log10(m) & $\text{ESA}_\kappa$ & 0.04 & 0.15 & 0.14 & 0 & 0.05 & 0.21 & \lt 0.09 & \lt 0.1 & \ze & \ze \\
\cmidrule{1-13}
\textbf{ESD}       & log10(\degree)          & $\text{ESD}_\mu$    & 1.17 & 1.03 & 1.4 & 0.9 & 1.03 & 1.43 & \lt 1.55 & \lt 1.6 & 2.228 & 1.08 \\
ESD STD            & log10(\degree)          & $\text{ESD}_\sigma$ & 0.07 & 0.1 & 0.08 & 0.3 & 0.08 & 0.06 & \lt 0.01 & \lt 0.17 & 0.30 & 0.36 \\
ESD decorr. dist.  & m                       & $\text{ESD}_\lambda$ & 9.2 & 12.5 & 10.4 & 12.4 & 6.4 & 11.6 & \lt 10.0 & \lt 20.0 & 4.0 & 4.0 \\
ESD freq. dep.     & log10(\degree)/log10(GHz) & $\text{ESD}_\gamma$ & \ze & \zel & \ze & \zel & \ze & \zel & \zeg & \zegl & -1.43 & \ze \\
ESD dist. dep.     & log10(\degree)/log10(m) & $\text{ESD}_\epsilon$ & \ze & 0.16 & -0.53 & -0.38 & -0.2 & -0.5 & \lt -0.5 & \lt -0.5 & \ze & \ze \\
ESD height dep.    & log10(\degree)/log10(m) & $\text{ESD}_\zeta$  & 0.13 & \ze & 0.3 & 0.54 & 0.26 & 0.25 & \lt 0.3 & \lt 0.14 & \ze & \ze \\
ESD STD freq. dep. & log10(\degree)/log10(GHz) & $\text{ESD}_\delta$ & \ze & \zel & \ze & \zel & \ze & \zel & \zeg & \zegl & 0.13 & \ze \\
ESD STD dist. dep. & log10(\degree)/log10(m) & $\text{ESD}_\kappa$ & \ze & -0.03 & 0.05 & -0.08 & 0.08 & 0.08 & \lt 0.09 & \lt 0.1 & \ze & \ze \\
\cmidrule{1-13}
\textbf{XPR}       & dB                      & $\text{XPR}_\mu$    & 13.0 & 12.8 & 18.0 & 15.2 & 16.1 & 15.0 & \lt 16.8 & \lt 14.4 & 11.0 & 10.0  \\
XPR STD            & dB                      & $\text{XPR}_\sigma$ & 1.6 & 1.3 & 2.6 & 1.8 & 3.2 & 2.1 & \lt 3.1 & \lt 2.4 & 4.0 & 4.0 \\
XPR decorr. dist.  & m                       & $\text{XPR}_\lambda$ & 16.5 & 7.1 & 23.4 & 6.9 & 14.9 & 16.3 & \lt 30.0 & \lt 27.0 & \ze & \ze \\
XPR height dep.    & dB/log10(m)             & $\text{XPR}_\zeta$ & -1.9 & -2.5 & -3.6 & 0.6 & -3.5 & -3.7 & \lt -4.5 & \lt -2.2 & \ze & \ze \\
\end{tabular}
\end{table*}

\subsection{Large-Scale Parameters}

\paragraph*{\Acf{PL}} The PL  depends on the 2D distance (i.e., the distance on the ground) between \ac{TX} and \ac{RX} and the carrier frequency. A dependence on the \ac{TX} height was not found. For both \ac{LOS} and \ac{NLOS}, results are close to the free space PL. For the \ac{NLOS} case, it means that there is strong coverage with high power despite the lack of a direct \ac{LOS} path. The 5.4~GHz measurements have 7~dB less power compared to the 2.37~GHz measurements. This is consistent with the theory of free-space propagation where the received power scales with the square of the frequency (i.e., 7.15~dB less power at 5.4~GHz compared to 2.37~GHz). Compared to 3GPP indoor office, there is a 6~dB increased LOS-PL in the industrial scenario\footnote{LOS @ 3.5 GHz, 50~m dist.: 78 dB for industrial, 72.6 dB for office }. This might come from the dimensions of the hall where reflected paths travel a longer distance compared to the office scenario. Also, there was a significant ground reflection in some measured trajectories which effected the average received power due to destructive interference. The NLOS-PL is very different from the 3GPP office model, where there is a 12~dB lower \ac{PL} in the measured industrial \ac{NLOS} scenario\footnote{\ac{NLOS} @ 3.5 GHz, 50~m dist.: 83.9 dB for industrial, 95.9 dB for office}. This indicates that the walls and objects in the factory reflect and scatter a significant portion of the transmitted signal.

\paragraph*{\Acf{SF}} \Ac{SF} occurs when an obstacle gets positioned between the \ac{TX} and the \ac{RX}. This leads to a reduction in signal strength because the wave is shadowed or blocked by the obstacle. This is modeled by log-normal distributed with three parameters: the reference \ac{STD} $\text{SF}_\sigma$ defines the width of the distribution, i.e., the power value (in dB) above or below the \ac{PL} at 1~GHz; the frequency dependence $\text{SF}_\delta$ describes how the \ac{SF} changes with frequency; and the decorrelation distance $\text{SF}_\lambda$ defines how fast the \ac{SF} varies when the terminal moves through the environment. The \ac{SF} values are significantly smaller compared to the 3GPP office environment. This also indicates that there might be good coverage due to a large number of reflected paths.

\paragraph*{\Acf{DS}} The DS is an important single measure for the delay time extent of a multipath radio channel. It is defined as the square root of the second central moment of the power-delay profile. As for the SF, the decorrelation distance $\text{DS}_\lambda$ defines how fast the \ac{DS} varies when the terminal moves through the environment. The \ac{DS} is calculated from both the delays $\tau$ and the path powers $P$, i.e., larger \acp{DS} can either be achieved by increasing the values of $\tau$ and keeping $P$ fixed or adjusting $P$ and keeping $\tau$ fixed. To avoid this ambiguity, an additional proportionality factor (delay factor) $r_\tau$ is introduced to scale the width of the distribution of $\tau$. In the measurements, the DS increases with increasing \ac{TX} height and increasing frequency. The measurements at 2.37~GHz have an average DS of around 30~ns whereas at 5.4~GHz, values around 80~ns are measured. These results are independent of the \ac{LOS} conditions. The 3GPP office scenario assumes values around 50~ns for \ac{NLOS} and 20~ns for LOS.

\paragraph*{\Acf{KF}} The \ac{KF} is the ratio between the power of the direct path and the power of further, scattered, paths\footnote{The KF can also be substantially large in \ac{NLOS} conditions, if there is a dominant specular path and diffuse scattered paths. This explains the results of the LOS-like path in the measurements.}. The \ac{KF} is assumed to be log-normal distributed defined by its mean value $\text{KF}_\mu$, its \ac{STD} $\text{KF}_\sigma$, the decorrelation distance $\text{KF}_\lambda$, and additional height and frequency-dependent terms. In the measurement data, the KF depends on the \ac{TX} height, where for increasing height, the KF decreases. This is consistent with the DS results: an elevated TX position leads to a larger number of scattered paths in the factory hall, and thus the relative strength of the direct path decreases. However, there are differences between the configurations. In case of Setup~1, at a TX height of 2~m, the \ac{KF} is only -3~dB\footnote{Setup 1: $\text{KF} = -1.6 - 5.7\cdot\log_{10}(2) \approx -3.3\ \text{dB}$}, which means that the direct path carries 1/3 of the total received power. In contrast, Setups 2\footnote{Setup 2: $\text{KF} = 4.8 - 8.2\cdot\log_{10}(2) \approx 2.3\ \text{dB}$} and 3\footnote{Setup 3: $\text{KF} = 2.8 - 3.6\cdot\log_{10}(2) \approx 1.7\ \text{dB}$} show different results at the same \ac{TX} height of 2~m. For both configurations, approximately 2/3 of the total received energy is carried by the \ac{LOS} component. In the \ac{NLOS} scenario there is a LOS-like path, i.e., a path that arrives within 5~ns of the expected \ac{LOS} arrival time that carries between 15\% and 50\% of the total received power. This explains the similar PL and DS results when comparing the \ac{LOS} and \ac{NLOS} measurements.

\paragraph*{\Acf{ASA}}
The \ac{ASA} does not depend on the distance or \ac{TX} height. Measured values vary between 10 and 90 degree in both \ac{LOS} and \ac{NLOS} conditions. This means that signals arrive at the receiver  from almost all directions. The averages observed can be found around 40\degree\ to 50\degree\ (5.4~GHz) and 40\degree\ (2.37~GHz). The 3GPP office shows similar values with a slightly smaller ASA for \ac{LOS} channels which might be due to the increased KF.

\paragraph*{\Acf{ASD}}
If both the \ac{TX} and the RX are at the same (2~m) height, the ASD is expected to have the same values as the ASA. For Setup~1, the values are around 35\degree , which is slightly smaller then the expected ASA value of 50\degree . This might be caused by the smaller TX antenna aperture of 205\degree\ which omits paths at the rear side of the antenna. Also, there is a dependence on the \ac{TX} height for \ac{NLOS} channels, leading to lower ASD values at higher \ac{TX} positions. For the other two setups, the measured ASD is much smaller at values below 5\degree\ for both configurations. This is due to the limited aperture of the transmit antennas. Hence, these low values have not been used for the combined industrial scenario. To obtain a consistent combination of the results for \ac{D2D} applications, the ASD at 2~m \ac{TX} height must be identical to the \ac{ASA} at 2~m \ac{RX} height. Hence, the combined results in Table~\ref{tab:all_parameters} take this into account by including the ASA results in the ASD evaluations at 2~m TX height.

\paragraph*{\Acf{ESA}}
The values of the ESA decrease with increasing distance between \ac{TX} and RX for all of the three antenna array configuration. This might be due to strong reflections at the ground and the roof of the factory hall. The incident angle at the reflector (i.e., at ground and roof) decreases with larger distances and so does the angle spread at the receiver. This is not considered by the 3GPP office scenario, but seems to be relevant for industrial systems.

\paragraph*{\Acf{ESD}}
The ESD depends on both the \ac{TX}-RX distance and the \ac{TX} height. For Setup~1, there is a height dependency for \ac{LOS} channels, where the elevation spread increases with increasing \ac{TX} height, but no distance dependency is observed. This effect does not occur in \ac{NLOS} channels where the ESD increases with increasing \ac{TX}-RX distance. At close \ac{TX}-RX distances, the results agree well with the ESA where both ESA and ESD have values around 15\degree . However, the ESD is much larger at large distances compared to the ESA. The results from Setup~2 increase with increasing \ac{TX} height and decrease with increasing distance. The distance dependency is consistent with the Setup~2 and 3 ESA results, but contradicts the Setup~1 results that show an increase in ESD with increasing distance. The contradictions in the results might be explained by the limited \ac{TX} antenna aperture. To obtain a consistent combination of the results for \ac{D2D} applications, the ESD at 2~m \ac{TX} height must be identical to the \ac{ESA} at 2~m \ac{RX} height. Hence, the combined results in Table~\ref{tab:all_parameters} take this into account by including the ESA results in the ESD evaluations at 2~m TX height.

\paragraph*{\Acf{XPR}}
The \ac{XPR} defines how the polarization changes for a multipath component, i.e., the initial polarization of a path is defined by the transmit antenna. However, for the \ac{NLOS} components, the transmitted signal undergoes diffraction, reflection or scattering before reaching the receiver. The \ac{XPR} (in dB) is assumed to be normal distributed. The cross polarization shows values between 13~dB and 16~dB for the 5.4~GHz measurements and a value of 13~dB in case of 2.37~GHz. The high values indicate that the  polarization remains almost unchanged for reflected waves. This is consistent with the 3GPP office scenario.

\subsection{Inter-Parameter Correlations}
The correlation values between the different parameters are listed in Table \ref{tab:Cross-Corr}. The upper right part (shown in white) contains the values for the \ac{LOS} channels, the lower left part shows the values for the \ac{NLOS} channels. The KF is negatively correlated with the DS, ASA, ASD, ESD and ESA and positively correlated with the SF and the XPR. The DS is positively correlated with the angular spreads. Azimuth and elevation spreads are positively correlated, both at the \ac{TX} and at the RX. The ESD is correlated with the ESA, whereas the ASD and the ASA are uncorrelated. This could mean that there is a specular reflection or a single scatterer either on the floor or on the ceiling of the factory hall that increases the elevation spread at both ends of the link simultaneously. But this does not affect the scattering in the azimuth plane.

\begin{table}[h]
   \setlength{\tabcolsep}{1.3pt}
    \scriptsize
    \centering
    \caption{Inter-Parameter Correlation Values}\label{tab:Cross-Corr}
    \vspace{-2mm}
\begin{tabular}{ lll | dddddddd }
 \multicolumn{3}{l|}{ Inter-Parameter } & \multicolumn{8}{c}{L O S} \\
 \multicolumn{3}{l|}{ Correlations  }
        &  \tbt{DS}
        &  \tbt{KF}
        &  \tbt{SF}
        &  \tbt{ASD}
        &  \tbt{ASA}
        &  \tbt{ESD}
        &  \tbt{ESA}
        &  \tbt{XPR} \\
 \hline \hline
 & DS  & 5.4 ODA &\tp 1    &  -0.57  & -0.3  & -0.14     & 0    & 0   & 0.32    & -0.18 \\
 &     & 5.4 Cross   &\tp 1      & -0.75   & -0.25   & 0.17   & -0.16   & 0.3  & 0.16  & -0.5  \\
 &	   & 2.4 Planar   &\tp 1      & -0.72  & -0.03   & 0.44   & 0.19   & 0.4  & 0.47  & -0.33  \\
 &     & \textbf{Combined} & \tp  1 &  \lt -0.7  &\lt  -0.3 & \lt  0.4  &\lt  0 &\lt  0.4 &\lt  0.3 &\lt   -0.4 \\
 &     & 3GPP Office & \tp  1 & -0.5 & -0.8 & 0.6 & 0.8 & 0.1 & 0.2 & \na \\
\cmidrule(l){2-11}

& KF   & 5.4 ODA  &\lo -0.72   &\tp 1      & 0.44    & -0.08     & 0   & 0.07     & -0.4   & 0.22  \\
 &     & 5.4 Cross    & \lo -0.46 &\tp 1   & 0.38    & -0.4   & 0.08   & -0.31      & -0.16  & 0.49  \\
 &     & 2.4 Planar   &\lo -0.69 &\tp 1   & 0.22    & -0.55   & -0.1   & -0.26      & -0.3  & 0.37  \\
 &     & \textbf{Combined}  &\lt  -0.6     &\tp 1  & \lt 0.4   &\lt -0.5  &\lt 0   &\lt -0.3 &\lt -0.3 &\lt  0.5 \\
 &     & 3GPP Office & \lona & \tp  1 &  0.5 & 0 & 0 & 0 & 0.1 & \na \\
\cmidrule(l){2-11}	

 & SF  & 5.4 ODA  &\lo 0.3  &\lo 0.39   &\tp 1      & 0.28  & 0.12   & 0.21      & 0.05  & 0.2  \\
 &     & 5.4 Cross    &\lo 0.03   & \lo 0.31 &\tp 1   & -0.05  & -0.22   & -0 .13     & -0.18  & 0.02 \\
 &     & 2.4 Planar    &\lo 0.39   & \lo -0.14 &\tp 1   & -0.1  & 0.03   & 0.05      & 0.1  & -0.12 \\
 &     & \textbf{Combined}   &\lt  0.4     &\lt 0  &\tp 1   &\lt 0  &\lt 0   &\lt 0 &\lt 0 &\lt  0 \\
 &     & 3GPP Office &\lo -0.5 & \lona & \tp  1 &  -0.4 & -0.5 & 0.2 & 0.3 & \na \\
\cmidrule(l){2-11}	

& ASD & 5.4 ODA  &\lo 0.07    &\lo -0.08      &\lo 0.11      &\tp 1     & -0.12    & 0.08    & 0.34  & 0.28 \\
N&     & 5.4 Cross    &\lo -0.41    & \lo 0.15 &\lo 0  &\tp 1  & -0.02    & 0.35    & 0.15  & -0.47  \\
L &	   & 2.4 Planar    &\lo 0.3    & \lo -0.22 &\lo 0  &\tp 1  & 0    & 0.3    & 0.26  & -0.41  \\
O &     & \textbf{Combined}   &\lt  0.2     &\lt 0  &\lt 0.2   &\tp 1  &\lt 0   &\lt 0.4 &\lt 0.2 &\lt  -0.5 \\
S &     & 3GPP Office &\lo 0.4 & \lona & \lo 0 & \tp  1 &  0.4 & 0.5 & 0 & \na \\
\cmidrule(l){2-11}

& ASA & 5.4 ODA  &\lo -0.35      &\lo 0.19   &\lo 0.14   &\lo 0   &\tp 1      & 0.15      & -0.13 & 0.07  \\
 &     & 5.4 Cross    &\lo 0.11    & \lo -0.04& \lo 0.44  &\lo -0.18 &\tp 1      & 0.16      & 0.33 & 0.26  \\
 &     & 2.4 Planar    &\lo 0.3    & \lo -0.03 &\lo 0.31  &\lo 0.06 &\tp 1      & 0.12      & 0.43 & -0.1  \\
 &     & \textbf{Combined}   &\lt  0     &\lt 0  &\lt 0.2   &\lt 0  &\tp 1   &\lt 0.1 &\lt 0.3 &\lt  0 \\
 &     & 3GPP Office &\lo 0 & \lona & \lo -0.4 & \lo 0  & \tp  1 &  0 & 0.5 & \na \\
\cmidrule(l){2-11}

 & ESD & 5.4 ODA  &\lo 0   &\lo -0.12      &\lo -0.21      &\lo 0   &\lo -0.07      &\tp 1      & 0.06  & 0.23  \\
&     & 5.4 Cross    &\lo 0   & \lo 0 &\lo 0.29   &\lo 0.04 &\lo 0.27      &\tp 1      & 0.52  & -0.24  \\
 &     & 2.4 Planar    &\lo 0.45   & \lo -0.34 &\lo 0.19   &\lo 0.35  &\lo 0      &\tp 1      & 0.31  & -0.04  \\
 &     & \textbf{Combined}   &\lt  0.3     &\lt -0.3  &\lt 0.3   &\lt 0.4  &\lt 0   &\tp 1 &\lt 0.3 &\lt  -0.2 \\
  &     & 3GPP Office &\lo -0.27 & \lona & \lo 0 & \lo 0.35 &\lo -0.08 & \tp  1 &  0 & \na \\
\cmidrule(l){2-11}

 & ESA & 5.4 ODA  &\lo 0   &\lo 0.04   &\lo 0.14   &\lo -0.19  &\lo 0.19   &\lo -0.1      &\tp 1  & -0.05   \\
 &     & 5.4 Cross    &\lo 0.03      &\lo 0.08  &\lo 0.37  &\lo -0.18 &\lo 0.31      &\lo 0.17      &\tp 1 & -0.1   \\
 &     & 2.4 Planar    &\lo 0.5      & \lo -0.26  &\lo 0.22  &\lo 0.05 &\lo 0.44      &\lo 0.19      &\tp 1 & -0.22   \\
 &     & \textbf{Combined}   &\lt  0.3   &\lt 0  &\lt 0.3   &\lt 0  &\lt 0.3   &\lt 0.3 & \tp 1 &\lt  -0.2 \\
 &     & 3GPP Office &\lo -0.06 & \lona & \lo 0 & \lo 0.23 &\lo 0.43 & \lo 0.42  &  \tp  1 &  \na \\
\cmidrule(l){2-11}

 & XPR & 5.4 ODA  &\lo -0.36   &\lo 0.26   &\lo 0   &\lo -0.08  &\lo 0.04    &\lo 0.04      &\lo 0  &\tp 1   \\
 &     & 5.4 Cross    &\lo 0.12      & \lo 0.02 &\lo 0.14  &\lo -0.16 &\lo 0.23      &\lo 0.06      &\lo 0.06 &\tp 1   \\
 &     & 2.4 Planar    &\lo -0.43      & \lo 0.27  &\lo -0.07  &\lo -0.27 &\lo -0.07      &\lo -0.32      &\lo -0.3 &\tp 1   \\
 &     & \textbf{Combined}   &\lt  -0.4     &\lt 0.3  &\lt 0   &\lt -0.2  &\lt 0   &\lt 0 & \lt -0.3 & \tp 1 \\
 &     & 3GPP Office & \lona &\lona &\lona &\lona &\lona &\lona &\lona &\tp  1\\
\end{tabular}
\vspace{-\baselineskip}
\end{table}

\section{Conclusions}

We performed several measurements under different typical industrial scenarios and conditions in various production halls with different topologies. The measurement data was processed and used to iteratively calculate the corresponding parameters that can be used in the 3GPP-NR channel model and the QuaDRIGa model. The outcome of our work is a propagations channel model that is able do describe radio propagation behavior in automation industry halls for typical industrial settings. In addition the model supports application cases like automatic guided vehicles and D2D communication in machine production halls. An open-source implementation of that model can be found at \cite{quadriga_www}.

\newpage
\section*{Acknowledgement}

The authors thank the Celtic Office and national funding authorities BMBF in Germany, Business Finland, and MINETAD in Spain for supporting this research and development through the ReICOvAir project. The project benefited also from the valuable technical contributions from GHMT AG, CETECOM GmbH, and Qosmotec GmbH in Germany; Trimek S.A. and SQS S.A. in Spain; Verkotan Ltd., Kaltio Technologies, and Sapotech in Finland. This research was also supported in part by the Academy of Finland 6Genesis Flagship (grant no. 318927).

\bibliographystyle{IEEEtran}
\bibliography{industrial_meas}

\end{document}